\documentclass[aps,pre,showpacs,floatfix,twocolumn,superscriptaddress]{revtex4-2}
\usepackage{graphicx}
\usepackage{dcolumn}
\usepackage{cancel}
\usepackage{bm}
\usepackage{tikz}
\usetikzlibrary{calc}
\usepackage{siunitx}
\usepackage{tablefootnote}
\usepackage{physics}
\usepackage[section]{placeins}
\definecolor{darkblue}{rgb}{0,0,.65}
\definecolor{darkgreen}{rgb}{1,0,0}
\usepackage{nicefrac}
\usepackage{booktabs}
\usepackage[%
  pdfauthor={Jinhong Zhu, Tao Chen, Zhiyi Li, Sheng Fang, and Youjin Deng},%
  pdfstartview=FitH,%
  breaklinks=true,%
  bookmarks=true,%
  colorlinks=true,%
  anchorcolor=black,%
  citecolor=red,
  filecolor=black,%
  menucolor=black,%
  urlcolor=blue,%
  linkcolor=red,%
 ]{hyperref}

\usepackage[english]{babel}
\usepackage{bm}
\usepackage{amssymb}
\usepackage{color, soul}
\usepackage{amsmath}
\usepackage{amstext}
\usepackage{latexsym}
\usepackage{graphicx}
\usepackage{mathtools}
\usepackage[ruled]{algorithm2e}
\usepackage{enumitem}

\newcommand{\scrA}{\mathcal{A}}

\newcommand{\scrG}{\mathcal{G}}

\newcommand{\scrS}{\mathcal{S}}
\newcommand{\scrC}{\mathcal{C}}
\newcommand{\scrN}{\mathcal{N}}

\newcommand{\scrT}{\mathcal{T}}

\begin{document}
\title{Percolation in the three-dimensional Ising model}
\author{Jinhong Zhu}%
\thanks{These two authors contributed equally to this paper.}
\affiliation{
Hefei National Laboratory for Physical Sciences at the Microscale and Department of Modern Physics, University of Science and Technology of China, Hefei 230026, China}

\author{Tao Chen}%
\thanks{These two authors contributed equally to this paper.}
\affiliation{
Hefei National Laboratory for Physical Sciences at the Microscale and Department of Modern Physics, University of Science and Technology of China, Hefei 230026, China}
\affiliation{Hefei National Laboratory, University of Science and Technology of China, Hefei 230088, China}

\author{Zhiyi Li}%
\thanks{peter0627@mail.ustc.edu.cn}
\affiliation{
Hefei National Laboratory for Physical Sciences at the Microscale and Department of Modern Physics, University of Science and Technology of China, Hefei 230026, China}
\affiliation{Hefei National Laboratory, University of Science and Technology of China, Hefei 230088, China}

\author{Sheng Fang}%
\thanks{shengfang9503@bnu.edu.cn}
\affiliation{School of Systems Science, Beijing Normal University, 100875 Beijing, China}

\author{Youjin Deng}%
\thanks{yjdeng@ustc.edu.cn}
\affiliation{
Hefei National Laboratory for Physical Sciences at the Microscale and Department of Modern Physics, University of Science and Technology of China, Hefei 230026, China}
\affiliation{Hefei National Laboratory, University of Science and Technology of China, Hefei 230088, China}

\begin{abstract}
Geometric representations provide a useful perspective on critical phenomena in the Ising model. In a recent study [\href{https://journals.aps.org/pre/abstract/10.1103/fqwr-ckj9}{Phys. Rev. E \textbf{112}, 034118 (2025)}], we found that the two-dimensional critical Ising model exhibits two consecutive percolation transitions for geometric spin clusters as the bond-occupation probability $p$ between parallel spins increases.
Here, through extensive Monte Carlo simulations, we show that this phenomenon does not persist in three dimensions, where we observe only a single percolation transition on critical Ising configurations. Further theoretical analysis of the Ising model on the complete graph also yields the same scenario. 
In addition, we study percolation on a two-dimensional layer embedded in the three-dimensional critical Ising model. For this layer system, we estimate the red-bond exponent $y_p=0.426(6)$ and the fractal dimensions of the largest cluster, hull, and shortest path as $d_f=1.892\,6(20)$, $d_{\rm hull}=1.663(4)$, and $d_{\rm min}=1.080(10)$, respectively. These values indicate a distinct universality class induced by coupling to out-of-plane critical correlations.
\end{abstract}
\maketitle

\section{Introduction}
\label{sec:intro}

The Ising model, originally introduced by Lenz and Ising~\cite{Ising1925}, is a prototypical lattice model in statistical physics and has long served as a central framework for understanding phase transitions and critical phenomena~\cite{duminil2022100}. 
It describes a system of binary spins $s_i=\pm 1$ located on the sites of a lattice. In the absence of an external magnetic field, its partition function is given by
\begin{equation}
    Z_{\rm spin} = \sum_{\{s_i\}} e^{K\sum_{\langle ij \rangle} s_i s_j},
\end{equation}
where $K$ denotes the reduced coupling strength, and the summations run over all possible spin configurations $\{s_i\}$ and all nearest-neighbor pairs $\langle ij \rangle$. 
Despite its simple formulation, the model exhibits strongly dimension-dependent behavior. In one dimension, thermal fluctuations destroy long-range order at any finite temperature~\cite{Ising1925}. In two dimensions, Onsager's exact solution provided the first rigorous demonstration of a continuous phase transition and laid the foundation for the modern theory of critical phenomena~\cite{Onsager1944}. In four dimensions and above, the critical behavior is governed by mean-field theory, with the correlation-length exponent $\nu = 1/2$ and the anomalous dimension $\eta = 0$.

In contrast, the three-dimensional (3D) Ising model lacks an exact analytical solution~\cite{aizenman1982geometric,baxter2016exactly}. This difficulty is commonly attributed to the absence of algebraic structures such as commuting transfer matrices and the Yang--Baxter equation, which make the two-dimensional model exactly solvable. As a result, the 3D Ising model has become a benchmark for numerical and field-theoretic studies of criticality. High-precision Monte Carlo (MC) simulations yield accurate estimates such as the critical point $K_c = 0.221\,654\,626(5)$ and the exponent $\nu = 0.629\,912(86)$~\cite{ferrenberg2018pushing,hou2019geometric}. On the analytical side, renormalization-group (RG) methods provide systematic estimates through $\epsilon-$expansion and field theory~\cite{WilsonKogut1974,zinn2007phase}, while conformal-bootstrap calculations now achieve even higher precision, for example $\nu = 0.629\,970\,97(12)$~\cite{chang2025bootstrapping,El2012Solving,Poland2019Conformal}. In parallel, Hamiltonian approaches such as fuzzy-sphere regularization provide a direct route to lattice realizations of critical conformal field theories~\cite{zhu2023uncovering,hu2023operator}. Together these approaches characterize the 3D Ising universality class with high precision, even though an exact solution remains unknown.

\begin{figure*}[t]
    \centering
    \includegraphics[width=\textwidth]{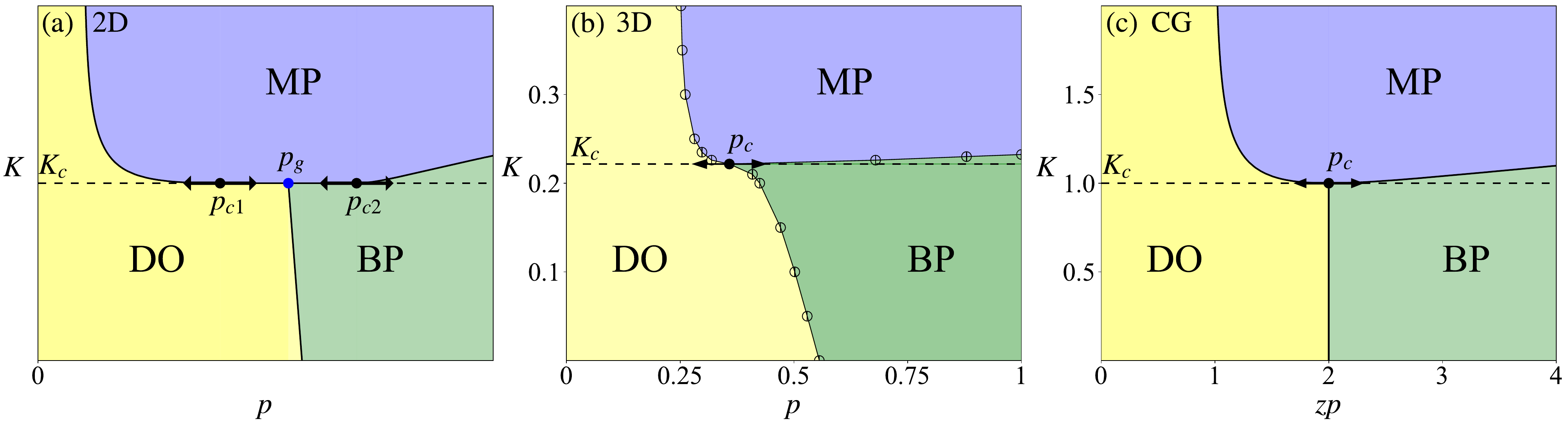} 

    \caption{The phase diagrams for percolation in the Ising model on the (a) square lattice (2D), (b) cubic lattice (3D), and (c) complete graph (CG). Different phases are indicated by various colors: the disordered phase (DO, yellow), the majority-spin percolated phase (MP, purple), and the phase where both majority- and minority-spin clusters percolate (BP, green). Panel (a) shows a pair of critical points $p_{c1}$ and $p_{c2}$ with a stable fixed point $p_{g}$ in between for percolation in the 2D Ising model. In higher dimensions, only a single phase transition point is observed on the critical line $K=K_{c}$, as shown in (b) and (c). The arrows indicate the direction of renormalization group flows.}
\label{fig:D_Phase}
\end{figure*}

In addition to the spin representation, the Ising model can be studied through geometric constructions. A standard example is the Fortuin-Kasteleyn (FK) representation~\cite{FortuinKasteleyn1972}, or random-cluster (RC) model with $q=2$~\cite{grimmett2006random}, whose partition function reads
    \begin{equation}
        Z_{\rm{RC}} =  \sum_{\scrA \subseteq \scrG} p^{b(\scrA)}(1-p)^{E-b(\scrA)} q^{k(\scrA)}. 
    \end{equation}
Here, the sum runs over all subgraphs $\scrA$ of the lattice $\scrG$. The quantity $b(\scrA)$ is the number of occupied bonds in $\scrG$, $E$ is the total number of edges in $\scrG$, and $k(\scrA)$ is the number of connected components. The probability $p$ is related to the Ising coupling by $p = 1 - e^{-2K}$, and the bond weight is $v = p/(1-p)$. The model reduces to standard bond percolation when $q=1$. 
This geometric perspective is not merely a representational change.
On one hand, it underpins the development of cluster MC algorithms, such as the Swendsen-Wang (SW) algorithm~\cite{SwendsenWang1987}, which dramatically reduce critical slowing down. 
On the other hand, it yields a much richer array of phenomena due to its broader range of geometric observables, many of which lack corresponding analogs in the spin representation. 
In two dimensions, the FK representation connects critical cluster interfaces to conformal field theory and Schramm-Loewner evolution (SLE)~\cite{cardy2005sle}. In higher dimensions it reveals distinct upper critical dimensions: the thermal upper critical dimension $d_c = 4$ and a separate geometric upper critical dimension $d_p = 6$ associated with cluster properties~\cite{Fang2022Geometric,FangGeometric2023}. Even on the complete graph (CG, $d \to \infty$), where all vertices are mutually connected and the free-energy density is analytically tractable, the FK representation retains nontrivial geometric structure~\cite{fang2021percolation}.

\begin{figure*}[t]
    \centering
    \begin{tikzpicture}[baseline=(current bounding box.center)]

    \node[inner sep=0pt] (img) {\includegraphics[width=\textwidth]{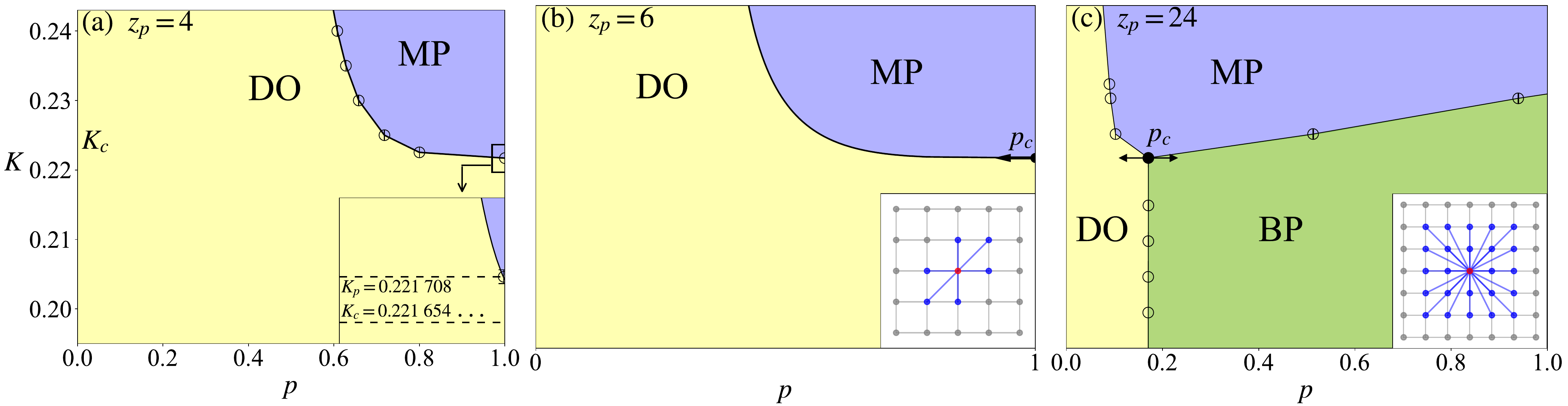}};
    \draw[dashed, black, line width=0.5pt] 
        ([xshift=0.95cm, yshift=2.84cm] img.south west) --
        ([xshift=-0.28cm, yshift=2.84cm] img.south east);
    \end{tikzpicture}
    \caption{The phase diagrams for the 3D Ising model with percolation coordination numbers (a) $z_p = 4$, (b) $z_p = 6$, and (c) $z_p = 24$.  The horizontal dashed line represents the critical line $K = K_c$. 
    The inset in (a) zooms in the phase diagram near $(p=1, K=K_{c})$, highlighting the deviation between $K_{p}=0.221\,708$ and $K_c= 0.221\,654\cdots$ for $z_{p}=4$. The insets in (b) and (c) show the equivalent neighbors for $z_p=6$ and $z_p=24$, respectively. For each inset, a given lattice site (red dot) connects to its neighbors (blue dots) through the edges (blue lines). }
\label{fig:Phase_layer}
\end{figure*}

Beyond the strict FK formulation, one can generalize percolation in the Ising model in two ways~\cite{chen2025percolation}. First, the constraint $p=1-e^{-2K}$ is relaxed, such that the bond-occupation probability $p$ becomes an independent control parameter. The limit $p=1$ corresponds to geometric-spin (GS) clusters. Second, bonds between parallel spins need not be restricted to nearest neighbors.
Using the equivalent-neighbor method~\cite{ouyang2018equivalent}, the percolation coordination number $z_p$ can be extended over a finite range, independent of the Ising coordination number $z_i$. Applied to the 2D critical Ising model, this framework reveals an unusual phase diagram. As the bond probability $p$ increases, the system undergoes two consecutive percolation transitions associated with the majority- and minority-spin clusters, respectively~\cite{chen2025percolation}. As shown again in Fig.~\ref{fig:D_Phase}(a), the phase diagram is then divided into a disordered phase (DO), a majority-percolated phase (MP), and a phase in which both majority- and minority-spin clusters percolate (BP). These regimes are separated by two thresholds, $p_{c1}$ and $p_{c2}$. The first transition is consistent with the FK-Ising universality class, whereas the second does not match any established universality class.

A natural question is whether these consecutive geometric transitions persist in higher dimensions. This issue is particularly intriguing given that the 3D bulk spin cluster geometry is already known to be unconventional. For instance, previous work found that, at criticality, the size of 3D FK clusters is asymptotically proportional to their boundary size.~\cite{Picco1993}. This stands in stark contrast to the two-dimensional case, where the cluster area and perimeter are both fractal objects characterized by distinct fractal dimensions.
Motivated by this broader geometric context, we study percolation in the 3D Ising model at the critical coupling $K_c= 0.221\,654\,631(8)$~\cite{hou2019geometric}. In contrast to the 2D case, we observe only a single percolation transition as $p$ increases from $0$ to $1$, irrespective of whether $z_i =z_p$. In the high-temperature region ($K < K_c$), the system crosses directly from the DO phase to the BP phase as $p$ increases, with the critical percolation threshold $p_c$. 
Namely, both the majority- and miority-spin clusters percolate simultaneously at $p_c$.
In the low-temperature region ($K>K_c$), it passes from the DO phase to the MP phase and then to the BP phase, with thresholds $p_{c1}$ and $p_{c2}$ for the majority- and minority-spin clusters, respectively. The estimates of $p_c$, $p_{c1}$, and $p_{c2}$ are listed in Table~\ref{tab:pc1pc2} in the Appendix~\ref{sec:AppendixA} and shown as black dots in Fig.~\ref{fig:D_Phase}(b), and, as long as $K \neq K_c$, the transitions are simply in the 3D uncorrelated percolation universality class. We find the same single-transition behavior at criticality on the CG through theoretical analysis, where details are shown in Appendix~\ref{sec:AppendixB}. Taken together, these results suggest that geometric spin clusters exhibit only one critical percolation transition for $d >2$.

Furthermore, we also study a 2D layer of the cubic lattice. This plane is coupled to the surrounding 3D bulk through the adjacent layers, and the spin-spin correlation function in the layer $g({\bf x})$ decays algebraically as $\|{\bf x}\|^{2-d-\eta}$ with $\eta= 0.036\,297\,612(48)$ for the 3D Ising model~\cite{chang2025bootstrapping}. Thus, it is expected that its percolation behavior should differ from that of ordinary 2D percolation in Ref.~\cite{chen2025percolation}. Previous work~\cite{SlicePicco2014} on the sliced layer of the 3D (SL3D) Ising model mainly concerned the case $(K=K_c, p=1)$: spin clusters on the layer were reported to be critical, with critical exponents different from those of the 2D critical Ising model. Here, by contrast, we study the more general case for percolation on the layer in the full $(K,p)$ plane. 
We determine the phase diagram in the $(K,p)$ plane for several percolation ranges $z_p$. As illustrated in Fig.~\ref{fig:Phase_layer}, the Ising spins on the layer have coordination number $z_i=4$, while the percolation coordination number takes the values $z_p=4$, $6$, and $24$. For $z_p = z_i = 4$, we find no phase transition along the critical line $K=K_c$ for $p\leq 1$, which differs from the observation in Ref.~\cite{SlicePicco2014}. Along the vertical line $p=1$, the MP phase sets in at $K_{p,c} = 0.221\,708(8)$, which is close to but clearly above the bulk 3D Ising critical coupling $K_c= 0.221\,654\,631(8)$~\cite{hou2019geometric}, as shown in Fig.~\ref{fig:Phase_layer}(a2).

\begin{table}
  \centering
  \caption{Comparison of critical exponents between the percolation transition on the SL3D Ising model and standard 2D percolation transition.}
\begin{ruledtabular}
  \begin{tabular}{ l l l l l }
    & $y_{p}$ & $y_{h}$ & $d_{\text{hull}}$ & $d_{\text{min}}$ \\
    \hline
    SL3D Ising & 0.426(6) & 1.892~6(20) & 1.663(4) & 1.080(10) \\
    \\
    2D Per. & 3/4 & 91/48 & 7/4 & 1.130~77(2)~\cite{PhysRevE.86.061101} \\
  \end{tabular}
\end{ruledtabular}
  \label{tab:exponents}
\end{table}

We then increase the percolation range. For $z_p=6$, the percolation process is analogous to site percolation on a triangular lattice, where the self-matching property yields that the critical threshold is located at $p_c=1$. 
At criticality, the scaling of the largest cluster gives the magnetic exponent $y_h=1.892\,6(20)$. We further estimate the hull and shortest-path exponents as $d_{\rm hull} = 1.663(4)$ and $d_{\rm min} = 1.080(10)$, and the RG exponent along the $p$ axis as $y_p = 0.426(6)$. These estimates, summarized in Table~\ref{tab:exponents}, differ significantly from the standard 2D percolation values and therefore contrast with earlier studies that related the critical behavior to established 2D universality classes~\cite{Saberi_2010,Saberi_2019}. 

For $z_p=24$, we obtain the percolation threshold $p_c=0.170\,57(13) < 1$ and obtain consistent estimates of critical exponents.
In addition, for $K > K_c$ we uncover two distinct thresholds associated with the percolation of majority- and minority-spin clusters, as shown in Fig.~\ref{fig:Phase_layer}(c).

The remainder of this paper is organized as follows. In Sec.~\ref{sec:simusamp}, we detail our simulation methodology. Numerical results are presented in Sec.~\ref{sec:numres}, and we conclude with a discussion in Sec.~\ref{sec:discussion}.

\section{Simulation and sampled quantities}
\label{sec:simusamp} 

We simulate the three-dimensional Ising model with periodic boundary conditions using the SW cluster algorithm for system sizes $L$ ranging from 16 to 128. For each size, we generate at least $10^6$ samples for statistical averaging. To determine the percolation phase diagram efficiently away from the magnetic critical coupling $K_c$, we use the event-based method~\cite{Fan2020gap, LiMing23, fang2024universal, Shi2025}. The basic idea is to identify characteristic events in a finite system, such as the maxima of the second-largest cluster size or susceptibility, or the largest increment in the size of the largest cluster. The locations of these events define finite-size pseudocritical points, which converge to the thermodynamic critical threshold as $L \to \infty$.

In practice, we occupy candidate bonds between parallel spins one by one. During a single bond-placement process, we record the gap in the largest cluster at step $\scrT$,
\begin{equation}
\Delta(\scrT) =\mathcal{C}_{1}(\scrT)-\mathcal{C}_{1}(\scrT-1).
\end{equation}
This quantity reaches its maximum at step $\scrT_{\text{max}}$. The ensemble average of $\scrT_{\text{max}}$ gives the pseudocritical threshold $p_{c}(L)=\langle \scrT_{\text{max}}/\scrN\rangle$, where $\scrN$ is the total number of available bonds and depends on the underlying Ising configuration. The finite-size scaling (FSS) form of $p_{c}(L)$ is
\begin{equation}
p_{c}(L)=p_{c}(\infty) + aL^{-y_p},
\label{eq:event}
\end{equation}
where $p_{c}(\infty)$ is the exact percolation threshold in the thermodynamic limit, and $y_p$ is the RG exponent along the $p$ direction.

\subsection{Bulk quantities}
For each spin configuration, we place bonds between adjacent parallel spins to obtain the corresponding geometric cluster configuration. Within this configuration, each cluster is assigned a label, majority or minority, which depends on the sign of the cluster's spin value relative to the total magnetization~\cite{chen2025percolation}. For each configuration, we sample the second and fourth moments of the cluster sizes as:
\begin{equation}
   \scrS_{2,\alpha}^{b} = \sum_{i}(\scrC^b_{i,\alpha})^2, \qquad
   \scrS_{4,\alpha}^{b} = \sum_{i}(\scrC^b_{i,\alpha})^4, 
\end{equation}
where $\alpha=0$, $\mathrm{m}$, and $\mathrm{M}$ denote all spin clusters, the minority- and majority-spin clusters, respectively. We then calculate the corresponding Binder ratios
\begin{equation}
    Q^b_{s,\alpha} = \frac{ \langle \scrS^b_{2,\alpha} \rangle^2 }{ \langle  3(\scrS^b_{2,\alpha})^2 - 2\scrS^b_{4,\alpha} \rangle },
\end{equation}
with $Q_s^b \equiv Q_{s,0}^b$.

\subsection{Layer quantities}
In addition to the bulk, we investigate a 2D layer embedded within the 3D model. Due to translation invariance under PBC, the selection of this 2D layer is arbitrary. We place bonds with probability $p$ between neighboring sites on the layer, using coordination numbers $z_p=4, 6,$ and $24$ to construct the percolation clusters. Because a large $z_p$ significantly reduces the percolation threshold, placing bonds one by one becomes highly inefficient. To remedy this, we employ an accelerated bond-placement approach~\cite{PhysRevE.66.066110,PhysRevE.72.016126} that bypasses vacant bonds and directly advances to the next occupied edge. The spacing $i$ between neighboring occupied bonds is drawn according to:
\begin{equation}      
    i=1+\left\lfloor \ln(r)/\ln(1-p) \right\rfloor,\nonumber
\end{equation}
where $r \in (0,1]$ is a uniformly distributed random number, and $\left\lfloor\cdot \right\rfloor$ is the floor function. This algorithm reduces the number of visited edges from $z_pV/2$ to $pz_pV/2$, dramatically improving efficiency for small values of $p$. For each bond configuration on the 2D layer, we sample the following geometric quantities:

\begin{itemize}
    \item The wrapping probabilities $\mathcal{R}_0$ and $\mathcal{R}_2$: For a given configuration and considering all clusters, if at least one cluster wraps around the lattice in both $x$ and $y$ directions, we set $\mathcal{R}_2 = 1$; otherwise, $\mathcal{R}_2 = 0$. If no cluster wraps around the lattice in any direction, we set $\mathcal{R}_0 = 1$; otherwise, $\mathcal{R}_0 = 0$. We also define the indicators $\mathcal{R}_{\text{0M}}$ and $\mathcal{R}_{\text{2M}}$ for majority clusters, and $\mathcal{R}_{\text{0m}}$ and $\mathcal{R}_{\text{2m}}$ for minority clusters.     
    \item The size of the largest cluster $\mathcal{C}_1$. 
    \item The second and fourth moments of the cluster size distribution
    $\mathcal{S}_{2, \alpha}=\sum_{i}\mathcal{C}^{2}_{i,\alpha}$ and $\mathcal{S}_{4}=\sum_{i}\mathcal{C}^{4}_{i,\alpha}$, 
    where $\mathcal{C}_{i,\alpha}$ denotes the size of the $i$-th largest cluster with $\alpha=\mathrm{m}$ for minority, and $\alpha=\mathrm{M}$ for majority.
    \item The maximum hull length $\mathcal{L}_{\text{hull}}$. The hull is defined as the outermost closed loop that encloses the cluster, excluding internal boundaries from holes.
    \item The shortest-path distance between two sites within the same cluster ${\mathcal{S}}_{p}$. During the simulation, we use a breadth-first search (BFS) algorithm to identify all sites within a cluster; the shortest path $\mathcal{S}_{p}$ corresponds to the maximum depth reached during a single BFS execution.
\end{itemize}
\par  

Based on the observables above, we calculate the following ensemble-averaged quantities:
\begin{enumerate}[label=(\roman*)]
    \item The critical polynomial $P_{bh}$:
        \begin{equation}
            P_{bh} =\frac{1}{2}\langle \mathcal{R}_{\text{2M}}+\mathcal{R}_{\text{0M}}-\mathcal{R}_{\text{0m}}-\mathcal{R}_{\text{2m}}\rangle;
            \label{eq:Pbh}
        \end{equation}  
    \item The mean size of the largest cluster $C_{1}=\langle \mathcal{C}_{1} \rangle$, which scales as $C_{1} \sim L^{y_{h}}$ at the critical point.
    \item The mean length of the largest hull $L_{\text{hull}}=\langle \mathcal{L}_{\text{hull}} \rangle$, which scales as $L_{\text{hull}} \sim L^{d_{\text{hull}}}$ at the critical point.
    \item The mean shortest-path distance $s_{p} = \langle {{\mathcal S}}_{p} \rangle$, which scales as $s_{p} \sim L^{d_{\text{min}}}$ at the critical point.
    \item The Binder ratios for minority- and majority- spin
    \begin{equation}
        Q_{s,\alpha} = \frac{ \langle \scrS_{2,\alpha} \rangle^2 }{ \langle  3(\scrS_{2,\alpha})^2 - 2\scrS_{4,\alpha} \rangle },
    \end{equation}
\end{enumerate}

\section{Numerical results}
\label{sec:numres} 

In this section, we present the numerical results and extract critical thresholds and exponents by least-squares fits to the MC data. To reduce finite-size corrections, we exclude small lattices by imposing a lower cutoff $L \geq L_{\text{m}}$ and monitor the corresponding change in $\chi^2$. Our preferred fits use the smallest $L_{\text{m}}$ that yields an acceptable goodness of fit and for which increasing $L_{\text{m}}$ does not lower $\chi^2$ by substantially more than one unit per degree of freedom. In practice, we regard $\chi^2/\mathrm{DF} \approx 1$ as satisfactory. Systematic uncertainties are estimated by comparing several fitting ansatzes.

\subsection{Percolation in the bulk Ising model}
\label{sec:perisn}

We begin with percolation in the 3D bulk Ising model.
For the 2D critical Ising model, Ref.~\cite{chen2025percolation} found two consecutive percolation transitions along the critical line $K=K_c$ as $p$ increases. In Fig.~\ref{fig:D_Phase}(a), these thresholds are denoted by $p_{c1}$ and $p_{c2}$. The first transition belongs to the 2D FK-Ising universality class, whereas the second appears to belong to a different universality class. Between these unstable RG fixed points lies a stable fixed point $p_g$ associated with the percolation of GS clusters. For a percolation range $z_p=80$, the reported thresholds are $p_{c1}=0.020~327(5)$ and $p_{c2}=0.345~82(2)$~\cite{chen2025percolation}.

\begin{figure}[h] 
\centering 
\includegraphics[width=1\columnwidth]{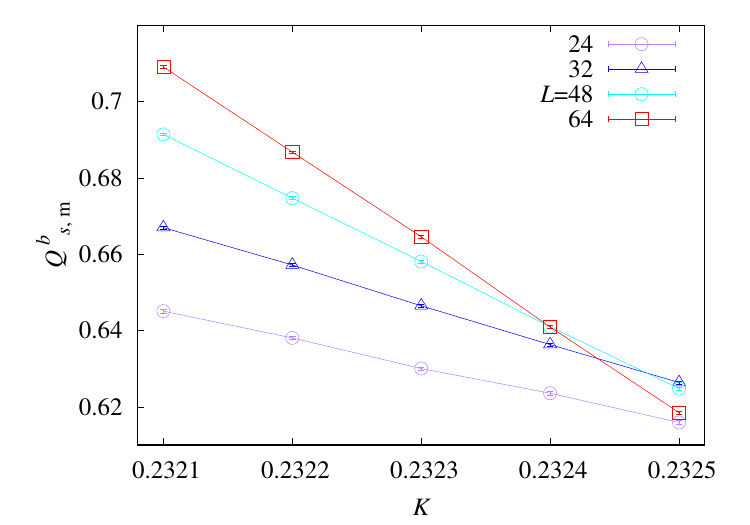}
\caption{Binder ratio $Q_{s,\mathrm{m}}^{b}$ for minority-spin clusters versus $K$ in the 3D bulk Ising model with the bond occupied probability $p=1$ for various system sizes $L$. The FSS analysis gives the estimate $K^{b}_{c}=0.232~38(9)$.} 
\label{fig:Qs2} 
\end{figure}

In contrast, for the 3D sing model on the critical line $K=K_c$, we find there is only a single percolation transition through extensive MC simulation results. 
As shown in Fig.~\ref{fig:D_Phase}(b), the percolation threshold locates  at $p_c=1-e^{-2K_c}$ for $z_p=z_i=6$, separating the DO and BP phases.
This suggests that both the majority- and minority-spin clusters percolate simultaneously. We further explore its phase diagram in the $(K,p)$ parameter space. In the high-temperature region ($K < K_c$), the system crosses directly from the DO phase to the BP phase as $p$ increases. However, in the low-temperature region ($K > K_c$), it crosses from the DO phase to the MP phase, and finally enters the BP phase. The phase boundary between the MP and BP phases, which corresponds to the percolation of the minority-spin clusters, terminates at $(K_p^b, p=1)$ with $K_p^b$ slightly larger than $K_c$. In Fig.~\ref{fig:Qs2}, we plot $Q_{s, \text{m}}^b$ versus $K$ along the vertical line $p=1$, where data from various system sizes intersect around $K \approx 0.232\,4$. We then perform a least-squares fit to the MC data of $Q_{s, \text{m}}$ using the general FSS ansatz:
\begin{equation}
    \label{Eq:fits}
\begin{aligned}
    \mathcal{O}(t,L)&=\mathcal{O}_c+\sum_{k=1}^m q_k (t L^{ y_p})^k + b_1 L^{y_1} + b_2 L^{y_2},
\end{aligned}
\end{equation}
where $\mathcal{O}_c$ is the value in the thermodynamic limit at criticality, $t$ is the deviation from criticality defined here as $t=K_p^b-K$, $m$ is the highest order of the scaling variable $tL^{y_{p}}$ with the corresponding scaling exponent $y_p$, and $q_k$ are the expansion coefficients. The terms $b_1L^{y_1}$ and $b_2L^{y_2}$ account for finite-size corrections with exponents $y_2 < y_1 < 0$. 
We first set $m=4$, $y_1=-1$, and $b_2=0$, which yields $K_c^b= 0.232\,38(9)$; in this fit, the coefficients $q_3$ and $q_4$ are statistically consistent with zero. Repeating the analysis with $m=2$ gives a compatible result. We also tested several other choices of the correction exponents $y_1$ and $y_2$, again obtaining mutually consistent estimates.
By comparing estimates from these various ansatzes, we ultimately obtain the estimate $K_c^b = 0.232\,38(9)$, which is in good agreement with previous results~\cite{MULLER1974,PhysRevE.91.042113}. \par  

We then extract the percolation thresholds at $K \neq K_c$. 
For majority spin part with $K>K_c$, the system can be treated as undergoing a 3D percolation transition with impurities. We adopt the event-based method to extract the critical thresholds
$p_{c1}$, which demonstrates superior efficiency and suffers from fewer finite-size corrections~\cite{LiMing23}. We perform a least-squares fit to the MC data using the ansatz in Eq.~\eqref{eq:event}, taking $y_p= 1.141\,3$, the known 3D critical percolation exponent~\cite{Xu2014}. While for the minority spin, the event-based approach is not applicable for determining the percolation threshold $p_{c2}$, we extract the percolation thresholds $p_{c2}(K)$ through an FSS analysis of the quantities $Q_{s, \text{m}}^b$.
These critical thresholds are summarized in Table~\ref{tab:pc1pc2} and plotted as black dots in Fig.~\ref{fig:D_Phase}(b).

The contrast between the 2D and 3D phase diagrams naturally raises the question of what happens in still higher dimensions. To address this point, we study the Ising model on the CG, where each site is connected to all others and the coordination number is $z = N-1$ for a system of size $N$. This graph can be regarded as the $d \to \infty$ limit of a hypercubic lattice. Its partition function can be derived analytically, and the derivation is given in Appendix~\ref{sec:AppendixB}. The resulting phase diagram is shown in Fig.~\ref{fig:D_Phase}(c). In the disordered phase ($K<K_{c}=1$), the two percolation thresholds coincide, giving a single threshold at $zp_c = 2$. In the ordered phase ($K>K_{c}=1$), broken $Z_{2}$ symmetry makes the majority- and minority-spin sectors inequivalent, so $p_{c1} \neq p_{c2}$. On the critical line $K=K_{c}=1$, however, there is again only a single threshold at $zp_{c}=2$.
Together with the numerical results in three dimensions, these analytical findings on the complete graph strongly suggest the conjecture that only a single percolation transition exists for geometric spin clusters in all spatial dimensions $d > 2$.

\begin{figure*}[t]
    \centering 
    \includegraphics[width=\textwidth]{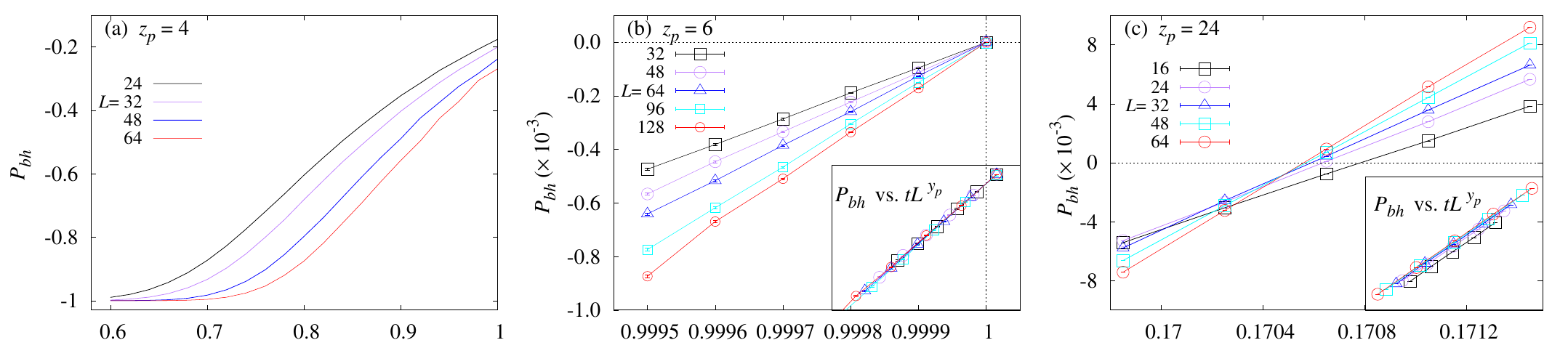} 
    \caption{Illustration of the percolation threshold $p_c$ on  SL3D Ising model for various percolation coordination numbers $z_p=$ (a) 4, (b) 6, and (c) 24 through the critical polynomial $P_{bh}$ versus bond probability $p$. 
    (a) For $z_p = 4$, curves for different system sizes $L$ do not intersect, indicating the absence of a percolation transition in the physical region $p < 1$. (b) For $z_p = 6$, the curves intersect sharply at $P_{bh} = 0$, locating the percolation threshold exactly at $p_c = 1$. The inset shows $P_{bh}$ vs. $tL^{y_p}$ with $y_p = 0.426$, where the excellent data collapse confirms the accuracy of the estimated exponent. (c) For $z_p = 24$, FSS yields a threshold $p_c = 0.170\,57(13) < 1$, with clear intersections observed among various system sizes.  The inset shows $P_{bh}$ vs. $tL^{y_p}$ with $y_p = 0.426$,
    the excellent data collapse confirms  the validity of the estimate of $y_p$ for both $z_p=6$ and $z_p=24$.}
\label{fig:Pbh_zp}
\end{figure*}

\begin{table*}
    \caption{Fitting results for $P_{bh}$ in percolation on the SL3D Ising model with $z_p = 6$ in the vicinity of $p = 1$ through the ansatz in Eq.~\eqref{Eq:fits}. Parameters without error bars were held fixed during the fitting procedure.}
    \begin{ruledtabular}
    \begin{tabular}[t]{l l l l l l l l l l}
    Obs. & $L_{m}$ & $\chi^{2}/DF$ & $y_{p}$ & $p_{c}$ & $\mathcal{O}_c$ & $q_{1}$ & $q_{2}$ & $b_{1}$ & $y_{1}$  \\
    \hline
    $P_{bh}$ & 24  & 29.6/21  & 0.426(8)& 1.000~004(6) & ~0.000~012(18)   &  -0.426(11) &  -3.8(20) & -0.000~3(3) & -1  \\
             & 24  & 29.2/21  & 0.426(8)& 1.000~002(5) & ~0.000~004(12)  &  -0.426(11) &  -3.8(20) & -0.000~5(3) & -2  \\
             & 24  & 25.7/21  & 0.422(8)& 1.000~000(4) & -0.000~006(8)  &  -0.428(11) &  -5.4(23) & ~0 & ~0  \\
             & 32  & 20.6/16  & 0.427(9)& 1.000~000(4) & -0.000~006(9)  &  -0.420(13) &  -4.9(22) & ~0 & ~0  \\
             & 24  & 28.9/23  & 0.424(3)& 1 & ~0 &  -0.434(6) &  -3.2(9) & ~0 & ~0  \\
             & 32  & 22.8/18  & 0.428(4)& 1 & ~0 &  -0.425(7) &  -3.0(8) & ~0 & ~0  \\
    \hline
    \end{tabular}
    \end{ruledtabular}
    \label{tab:yp}
\end{table*}
\subsection{Percolation in the layer of the 3D Ising model}

From our previous analysis, we found that the phase diagram in 2D fundamentally differs from those in higher dimensions, suggesting that the spatial dimension $d=2$ is special. We now consider a 2D layer of the 3D Ising model and investigate its percolation transitions, namely, the percolation transitions on the SL3D Ising model. 
In this layered setup, the geometrical clusters are constrained to the spatial dimension $d=2$, while the underlying critical spin fluctuations are governed by the 3D Ising model. Namely, the critical two-point spin correlation function behaves as $g({\bf x}) \sim \|{\bf x} \|^{2-d-\eta}$ with $\eta = 0.036\,297\,612(48)$, taking on the 3D exponent value~\cite{chang2025bootstrapping}. 

In previous studies, the properties of 2D layers and interfaces embedded within the 3D Ising model have been explored extensively. For the 2D boundaries of critical 3D clusters, topological analyses showed that the cluster volume scales linearly with surface area, indicating a branching instability of the boundary surfaces~\cite{Picco1993}. For a strictly planar layer, GS clusters were found to percolate at the 3D bulk Curie temperature, and their interfaces were argued to exhibit conformal invariance described by SLE~\cite{Saberi_2010}. The same setting also shows a ``super-roughening'' state at criticality, with local anomalous scaling characterized by geometric exponents matching those of the pure 2D Ising model~\cite{Saberi_2019}.

\subsubsection{Phase diagram}
\label{sec:layeredper}

Here, we examine the phase diagram for the layer system. The analysis parallels that used for the bulk model, and the resulting phase diagrams are shown in Fig.~\ref{fig:Phase_layer}. We begin with $z_p=4$, where the percolation bonds are restricted to nearest neighbors. In the low-temperature region ($K>K_c$), the system crosses from the DO phase to the MP phase; the corresponding thresholds $p_c$ are listed in Table~\ref{tab:pc1pc2}. As $K \to K_c$, the phase boundary terminates at $(p=1, K_{p,c})$, with $K_{p,c}$ slightly above the 3D Ising critical point $K_c = 0.221\,654\,631(8)$. Along the critical line $K=K_c$, Fig.~\ref{fig:Pbh_zp}(a) shows that the critical polynomial $P_{bh}$ does not exhibit an intersection for $p \in(0,1]$, indicating the absence of a phase transition in the physical interval and differing from the scenario proposed in Ref.~\cite{Saberi_2010}. Restricting to the vertical line $p=1$, the FSS analysis gives $K_{p,c}= 0.221\,708(8)$, again close to but clearly larger than $K_c$. This difference is highlighted in the inset of Fig.~\ref{fig:Phase_layer}(a).

We then increase the percolation range to $z_p=6$ and $24$, and the corresponding phase diagrams are shown in Figs.~\ref{fig:Phase_layer}(b) and (c), respectively. For $z_p=6$, the percolation process is effectively conducted on a triangular lattice, as illustrated in the inset of Fig.~\ref{fig:Phase_layer}(b). At $p_c=1$, any two neighboring spins with the same value belong to the same cluster, so the construction is analogous to site percolation on the triangular lattice, with one spin species interpreted as occupied and the other as empty. Since the self‑matching property of the triangular lattice determines the exact site-percolation threshold at $1/2$,  accordingly,  the line $p=1$ in the regime $K \le K_c$ is critical because the two spin species occur with equal density in the thermodynamic limit. In particular, at $K=K_c$ the percolation threshold is exactly $p_c=1$. Figure~\ref{fig:Pbh_zp}(b) provides numerical support: the curves of $P_{bh}$ intersect at $p=1$, and the crossing value is consistent with $0$, which can be derived through the self-matching property.
For the longer range $z_p=24$, FSS gives a threshold $p_c =0.170\,57(13) < 1$, as supported by the intersection of curves in Fig.~\ref{fig:Pbh_zp}(c). In this case, the BP phase appears on both sides of $K_c$.  We estimate $p_{c1}$ for $K>K_c$ with the event-based method. For $K<K_c$, $p_c$ is obtained from FSS analysis of the critical polynomial $P_{bh}$, while for $K>K_c$, $p_{c2}$ is estimated through FSS analysis of $Q_{s,\text{m}}$. The resulting thresholds are listed in Table~\ref{tab:pc1pc2} and shown as black dots in Fig.~\ref{fig:Phase_layer}(c).

\subsubsection{Critical exponents}
\label{sec:universalclass}
We now turn to the critical behavior near the fixed point $(K_c, p_c)$. Since no percolation transition occurs in the physical interval $p \leq 1$ along $K=K_c$ for $z_p=4$, we focus on $z_p=6$. The self-matching property is especially useful here: it fixes the threshold exactly at $p_c=1$ and implies $P_{bh}=0$ at criticality~\cite{chen2025percolation}. These exact constraints substantially improve the precision of the exponent estimates.

We fit the MC data for $P_{bh}$ with the ansatz in Eq.~\eqref{Eq:fits}, using the reduced variable $t = p_c-p$. When all parameters are left free, we obtain $y_{p}=0.426(9)$, while $p_c$ and $\mathcal{O}_c$ converge numerically to $1$ and $0$, respectively, as expected from self-matching property. The correction amplitude $b_1$ is also consistent with zero, which is compatible with the asymptotic behavior discussed in Ref.~\cite{Jacobsen_2012}. We therefore also perform constrained fits with $\mathcal{O}_c =0$, $p_c=1$, and $b_1=0$, obtaining $y_p = 0.424(3)$ for $L_m = 24$. Comparing several such ansatzes leads to the final estimate $y_{p}=0.426(6)$; the individual fits are summarized in Table~\ref{tab:yp}. The inset of Fig.~\ref{fig:Pbh_zp}(b) shows the collapse of $P_{bh}$ against $tL^{y_p}$ for $y_p=0.426$. 
In addition, in the inset of Fig.~\ref{fig:Pbh_zp}(c), we also plot the $P_{bh}$ versus the scaling variable $tL^{y_p}$, and the good data collapse suggests $z_p=24$ shares the same universality class with $z_p=6$. 
This exponent differs markedly from the standard 2D percolation value $y_{p}=3/4$.

\begin{figure*}[t]
    \centering 
    \includegraphics[width=1.0\textwidth]{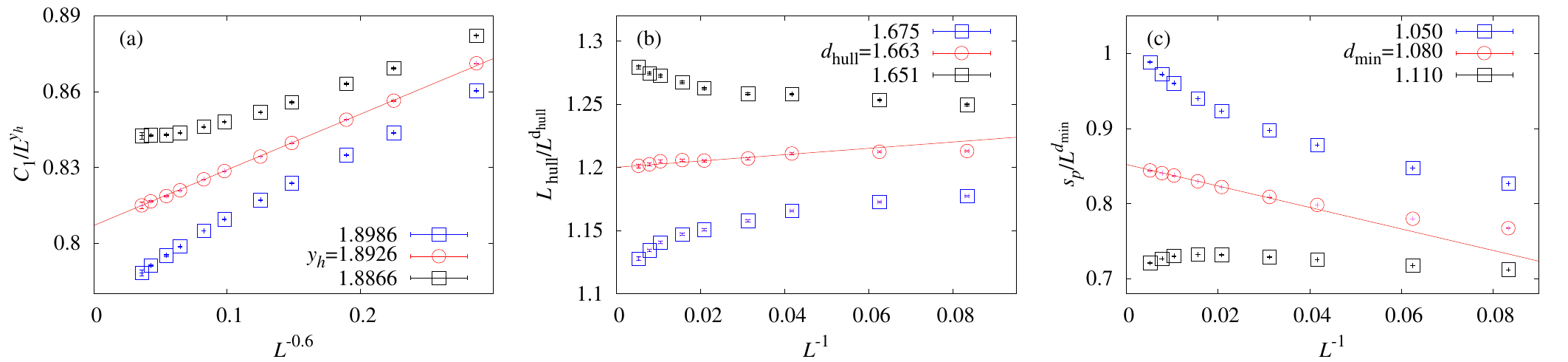} 
    \caption{Plots of $\mathcal{O}/L^{y_{\mathcal{O}}}$ versus $L^{y_1}$ for the critical SL3D Ising model with $z_p = 6$ at $p = p_c = 1$, with $y_1$ taking the correction exponent shown in Table~\ref{tab:yo}. 
(a) The largest cluster size $ C_1$ yielding the fractal dimension $y_h = 1.892\,6(20)$. 
(b) The hull length $ L_{\mathrm{hull}}$ with the estimated exponent $d_{\mathrm{hull}} = 1.663(4)$. 
(c) The shortest-path distance $s_p$ with the estimated exponent $d_{\mathrm{min}} = 1.080(10)$. 
The upper and lower curves indicate the scaling behavior when $y_{\mathcal{O}}$ is varied by $\pm 3$ standard errors from its central estimate.}
\label{fig:yo}
\end{figure*}

In addition to the RG exponent along the $p$ direction, we extract the magnetic exponent $y_h$, which coincides with the fractal dimension of the largest critical cluster $\mathcal{C}_1$. At criticality we fit the data with
\begin{equation}
    \label{Eq:fityh}
    \mathcal{O} =L^{y_{\mathcal{O}}}(a+b_{1}L^{y_{1}}+b_{2}L^{y_{2}}).
\end{equation}
Here $\mathcal{O}$ denotes the observable and $y_{\mathcal{O}}$ its scaling exponent. For $\mathcal{O}=C_1$, we identify $y_{\mathcal{O}}=y_h$. The terms $b_1L^{y_{1}}$ and $b_2L^{y_{2}}$ describe finite-size corrections, with $y_2 < y_1 < 0$. Setting $b_2 = 0$ and using $L_m = 16$ gives $y_h = 1.891(3)$ and $y_1 = -0.70(14)$. 
We then fix $y_1 = -0.6$ and obtain the more precise estimate $y_h = 1.892\,6(20)$. The fit details are listed in Table~\ref{tab:yo}. In Fig.~\ref{fig:yo}(a), we plot $C_1/L^{y_h}$ against $L^{-0.6}$. The near-linearity of the central curve, together with the opposite bending of the curves obtained by shifting $y_h$ by three standard errors, supports the quoted uncertainty. We note that this estimate lies close to the standard 2D percolation value $91/48 =1.895\,833\cdots$, so the two cannot be cleanly separated at the present numerical precision.

We also study two additional geometric exponents: the hull exponent $d_{\text{hull}}$ and the shortest-path exponent $d_{\text{min}}$. The corresponding MC data for $L_{\text{hull}}$ and $s_p$ are fitted with Eq.~\eqref{Eq:fityh}. For $L_{\text{hull}}$, fits with $b_{2}=0$ and all remaining parameters free are unstable, so we test several fixed values of the correction exponent $y_{1}$. The resulting estimates are mutually consistent, and combining them gives $d_{\text{hull}} = 1.663(4)$, clearly below the standard 2D value $7/4=1.75$. The fit details are summarized in Table~\ref{tab:yo}. 
Figure~\ref{fig:yo}(b) shows $L_{\text{hull}}/L^{d_{\text{hull}}}$ against $L^{-1}$ with $d_{\text{hull}}=1.663\pm0.012$ (three times the standard error). The near-linearity of the central curve supports the quoted uncertainty. We analyze the shortest path $s_p$ in the same spirit. 
Fixing $y_2=0$ results in large residuals in $\chi^2$ even for a large cutoff $L_m$, so we fix $y_2 =2y_1$. 
The results are listed in Table~\ref{tab:yo} and give the estimate $d_{\text{min}}=1.080(10)$, again distinct from the 2D percolation value $1.130\,77(2)$. Figure~\ref{fig:yo}(c) shows $s_{p}/L^{d_{\text{min}}}$ versus $L^{-1}$ for $d_{\text{min}}=1.050$, $1.080$, and $1.110$, and the nearly linear behavior for the central value supports the quoted estimate and uncertainty.

\section{Discussion}
\label{sec:discussion}

In this work, we investigated percolation of geometric spin clusters in the three-dimensional Ising model from both bulk and layered perspectives. For the bulk system, our Monte Carlo results show that, unlike the 2D critical Ising model, the 3D model exhibits only a single percolation transition at criticality. The same behavior is found for the complete graph, which supports the conjecture that geometric spin clusters have only one critical percolation transition for $d > 2$.

We also studied a 2D layer embedded in the 3D critical Ising model. For $z_p=4$, no percolation transition occurs along the physical interval of the critical line in the physical region $p \leq 1$. For longer percolation ranges, however, a transition emerges. In particular, for $z_p = 6$ the self-matching property fixes the threshold at $p_c = 1$ and implies $P_{bh}=0$ at criticality, enabling precise exponent estimates. Our FSS analysis gives $y_p = 0.426(6)$, $y_h = 1.892\,6(20)$, $d_{\text{hull}} = 1.663(4)$, and $d_{\text{min}} = 1.080(10)$.

These critical exponents differ significantly from those of standard 2D percolation, indicating that the layered case belongs to a distinct universality class shaped by the coupling to 3D critical correlations. More broadly, the results suggest that long-range correlations in the spin background can strongly modify geometric critical behavior. Clarifying how the percolation universality class is determined by the underlying spin correlations remains an interesting open problem.

\begin{table*}[!htbp]
  \centering
  \caption{Fitting results for geometric quantities in the SL3D Ising model with $z_p = 6$ and $p=p_c=1$ via the ansatz Eq.~\eqref{Eq:fityh}. Parameters without an error bar were held fixed during the fitting process.}
  \label{tab:yo}
  \setlength{\tabcolsep}{15pt}
  \begin{tabular}{l c c c c c c c}
    \toprule
    Obs. & $L_m$ & $\chi^2/\mathrm{DF}$ & $y_{\mathcal{O}}$ & $a$  & $b_1$ & $b_2$ & $y_1$ \\
    \midrule
   $C_{1}$ & 12  & 7.7/6  & 1.893(4)  & 0.80(2)  & \ 0.217(5) & 0 & \ \ \ \ \  -0.56(9) \\
   & 16  & 6.1/5  & 1.891(3)  & 0.82(2)  & 0.24(3)  & 0 & \ \ \ \ \ \ \  -0.70(14) \\
    & 24  & 5.6/4  & \ 1.896(12) & 0.78(8)  & 0.21(2)  & 0 & \ \ \ ~-0.5(2) \\
    & 16  & 6.7/6  & \ 1.8928(6) & \ 0.806(2) &\  0.224(7) & 0 & -0.6 \\
    & 24  & 5.7/5  & \ 1.8923(8) & \ 0.808(4) & 0.22(1)  & 0 & -0.6 \\
    \midrule
 $L_{\text{hull}}$     & 16  & 7.2/4  & 1.662(1)  & \  1.211(8)  & 0.11(6)  & 0 & -1 \\
    & 24  & 5.6/3  & 1.664(2)  & 1.21(1)   & \ \ 0.26(13) & 0 & -1 \\
    \midrule
 $s_{p}$     & 24  & 1.6/3  & 1.073(2)  & 0.89(1)   & \ -2.4(3)  & 16(3)& -1 \\
    & 48  & 1.0/2  & 1.077(2)  & 0.86(1)   & \ -1.6(2) & 0 & -1 \\
    \bottomrule
  \end{tabular}
\end{table*}

\section{ACKNOWLEDGMENTS}
This work was supported by the National Natural Science Foundation of China (under Grant No. 12275263 and No. 12505036), the Innovation Program for Quantum Science and Technology (under Grant No. 2021ZD0301900), and the Natural Science Foundation of Fujian Province 802 of China (under Grant No. 2023J02032). Y.D. acknowledges the valuable discussion with Abbas Saberi.

\appendix
\setcounter{section}{0} 
\setcounter{figure}{0} 
\setcounter{equation}{0} 
\setcounter{table}{0} 
\renewcommand{\thefigure}{A\arabic{figure}}
\renewcommand{\thetable}{A\arabic{table}}
\renewcommand{\theequation}{A\arabic{equation}}

\section{Data for Percolation Thresholds of the 3D Ising Model and 2D Layers}
\label{sec:AppendixA} 
This appendix presents the percolation thresholds for the 3D bulk Ising model and for the layered system with $z_p = 4$ and $z_p = 24$. Table~\ref{tab:pc1pc2} displays $p_c$ for $K \le K_c$ and the majority- and minority-spin thresholds $p_{c1}$ and $p_{c2}$ for $K > K_c$. A ``/'' entry indicates that the corresponding transition is absent in the physical range $p \in [0,1]$. These results support the phase diagrams discussed in the main text.

\section{Percolation in the Ising model on the complete graph}
\label{sec:AppendixB} 

The complete graph (CG) is a special network topology in which any two sites are directly adjacent. Consequently, the coordination number of any site is $z = N-1$ for a system with a total volume of $N$ sites. The CG can be formally regarded as the infinite-dimensional limit ($d \to \infty$) of a hypercubic lattice, where both the pure Ising model and its corresponding percolation models become analytically solvable. Here, we consider percolation on the CG Ising model, which can be decomposed into two distinct percolation processes occurring on subgraphs of sizes $N_M$ and $N_m$. 
The parameters $N_M$ and $N_m$ represent the total number of majority- and minority-spin sites, respectively, satisfying the volume constraint $N_M + N_m = N$. In the following derivation, we first analytically obtain expressions for $N_M$ and $N_m$, which subsequently allow us to determine the respective critical percolation thresholds $p_{c1}= 1/N_M$ and $p_{c2}=1/N_m$. \par  

\begin{table*}[!t]
\caption{Percolation thresholds for the 3D Ising model and SL3D Ising model with $z_{p}=4$ and $z_p=24$. In the disordered phase ($K \leq K_c$), a single threshold $p_c$ is observed, and for $z_p=4$, the absence of a value for $p_{c}$ (indicated by ``/'') signifies that the system is in the DO phase within the physical range $p \in [0,1]$. In the ordered phase ($K > K_c$), the percolation transition line splits into $p_{c1}(K)$ and $p_{c2}(K)$, corresponding to the thresholds for the majority- and minority-spin clusters, respectively. The absence of a value for $p_{c2}$ (indicated by ``/'') signifies that the minority-spin component does not percolate, meaning $p_{c2}$ is located in the non-physical regime $p>1$.}
\begin{ruledtabular}
\begin{tabular}{c c c c c c c c}

&  & $K$  & $p_{c}$  &  & $K$  &  $p_{c1}$  & $p_{c2}$  \\
\hline
3D  & $K\leq K_{c}$  &   0.00 & 0.555~97(2)  & $K>K_{c}$  & 0.226 & 0.319~04(4)  & 0.680~1(3)    \\
    &            &  0.05 & 0.529~44(7)   &           & 0.230 & 0.307~46(2)  & 0.878~7(3)    \\
    &            &  0.10 & 0.502~01(4)   &            & 0.235 & 0.297~95(3)  &   /           \\
    &            &  0.15 & 0.470~93(3)   &            & 0.250 & 0.281~62(3)  &   /          \\
    &            &  0.20 & 0.424~87(4)   &            & 0.300 & 0.261~14(2)  &   /          \\
    &            &  0.21 & 0.408~92(4)   &            & 0.350 & 0.254~47(2)  &   /      \\
    &            & $K_{c}$ & $1-e^{-2K_{c}}$  &        & 0.400 & 0.251~59(3)  &   /      \\ 
\hline
 2D layer  & $K\leq K_{c}$  &  / & /    & $K>K_{c}$  & 0.225 & 0.718~2(2)  &  /   \\
  $z_p=4$  &            &  / & /    &            & 0.230 &  0.658~0(8) & /    \\
    &            &  / & /    &            & 0.235 & 0.628~0(2)  & /    \\
   &            &  / & /    &            & 0.240 & 0.607~8(1)  & /    \\
\hline
 2D layer  & $K\leq K_{c}$  &  0.200 & 0.170~30(6)    & $K>K_{c}$ & 0.225 & 0.102~4(2)  & 0.520~9(3)    \\
 $z_p=24$   &            &  0.205 & ~\ 0.170~26(10)    &           & 0.230 & 0.092~0(2)  &\, 0.926~8(12)     \\
    &            &  0.210 & 0.170~52(3)    &                & 0.232 & 0.089~6(1)  &   /     \\
    &            &  0.215 & ~\ 0.170~70(17)    &              & &   &        \\
    &            &  $K_{c}$ & ~\  0.170~57(13)  &            & &   &           \\

\end{tabular}
\end{ruledtabular}
\label{tab:pc1pc2}
\end{table*}

To systematically calculate the spontaneous macroscopic magnetization $M = \sum_{i=1}^N \sigma_i$, we introduce a uniform external magnetic field $h$. The Hamiltonian for the CG Ising model in the presence of this field is given by
\begin{equation}
\mathcal{H}/k_{B}T = -\frac{K}{N}\sum_{i \neq j} \sigma_{i}\sigma_{j} - h \sum_{i=1}^N \sigma_i,
\end{equation}
where $K$ is the coupling constant, $h$ is the dimensionless external field, and the sum runs over all pairs of adjacent sites. The prefactor $1/N$ ensures the extensivity of the total energy. Noting that $\sum_{i \neq j} \sigma_i \sigma_j = M^2 - N$, the partition function can be expressed as
\begin{equation}
    Z(K,h) = \sum_{\{\sigma_i\}} e^{-K} e^{\frac{K}{N} M^2 + h M}.
\end{equation}
To decouple the quadratic spin interaction, we apply the Hubbard-Stratonovich transformation
\begin{align}
& \exp\left( \frac{K}{N} M^2 \right) = \sqrt{\frac{N}{4\pi K}} \int_{-\infty}^{\infty} dx \, \exp\left( -\frac{N}{4K} x^2 + x M \right).
\end{align}
Substituting this into the partition function allows us to perform the exact summation over the spin configurations $\{\sigma_i\}$
\begin{align}
    Z(K,h) &= e^{-K} \sqrt{\frac{N}{4\pi K}} \int_{-\infty}^{\infty} dx \, e^{-\frac{N x^2}{4K}} \sum_{\{\sigma_i\}} e^{(x+h)M} \nonumber \\
           &= e^{-K} \sqrt{\frac{N}{4\pi K}} \int_{-\infty}^{\infty} dx \, e^{-\frac{N x^2}{4K}} \left[ \sum_{\sigma=\pm 1} e^{(x+h) \sigma} \right]^N \nonumber 
\end{align}
In the large-$N$ approximation, the free energy per spin $f(x,h) = -\frac{1}{N} \ln Z(K,h)$ takes the form
\begin{equation}
    f(x,h) \approx \frac{x^2}{4K} - \ln \left(2\cosh (x+h) \right).
\end{equation}
In the thermodynamic limit ($N \to \infty$), the integral is overwhelmingly dominated by the saddle point $x^*$ that minimizes $f(x,h)$. The saddle-point condition without an external  field is thus given by
\begin{equation}
\left. \frac{\partial f(x,h)}{\partial x} \right |_{h=0} = \frac{x^*}{2K} - \tanh(x^*) = 0.
\end{equation}
The spontaneous magnetization density $\langle m \rangle$ is obtained by taking the first derivative of the free energy with respect to the external field $h$
\begin{equation}
\left. \langle m \rangle = -\frac{\partial f(x^*,h)}{\partial h}\right|_{h=0} = \tanh(x^*).
\end{equation}
We then obtain the self-consistency equation
\begin{equation}
  \label{eq:m}
  \langle m \rangle = \tanh(2K \langle m \rangle).
\end{equation}
Note that the magnetization density is given by $m= \frac{1}{N}(N_M -N_m)$, such that the ensemble-averaged numbers of majority- and minority-spin sites are determined by:
\begin{equation}
     \langle N_M \rangle = N(1+\langle m \rangle)/2, \quad \langle N_m \rangle  = N(1-\langle m \rangle)/2.
\end{equation}
Thus, we obtain the renormalized critical percolation thresholds for the majority-spin $zp_{c1}$ and minority-spin $zp_{c2}$ components on the CG as
\begin{align}
zp_{c1} &= \frac{N}{\langle N_M \rangle} = \frac{2}{1+\langle m \rangle}, \\
zp_{c2} &= \frac{N}{\langle N_m \rangle} = \frac{2}{1-\langle m \rangle},
\end{align}
where $\langle m \rangle$ is the solution to Eq.~\eqref{eq:m}.

\bibliography{references.bib}
\end{document}